\documentstyle[aabib,epsfig]{l-aa-ps}

    \newcommand{\nh}{\mbox{$ N({\rm H}) $}}
    \newcommand{\cq}{\mbox{$ \chi^2$}}

\begin{document}
%\draft
\thesaurus{09.01.1; 09.09.1 Cas~A; 09.19.2; 13.25.4}

\title{The broad-band X-ray spectrum of the Cas~A supernova remnant as
  seen by the BeppoSAX observatory}

\author{F. Favata\inst{1} \and J. Vink\inst{2} \and D.
  Dal~Fiume\inst{3} \and A.\,N. Parmar\inst{1} \and A.
  Santangelo\inst{4} \and T. Mineo\inst{4} \and A.
  Preite-Martinez\inst{5} \and J.\,S. Kaastra\inst{2} \and J.\,A.\,M.
  Bleeker\inst{2}}

\institute{Astrophysics Division -- Space Science Department of ESA, ESTEC,
 Postbus 299, NL-2200 AG Noordwijk, The Netherlands 
\and
SRON Laboratory for Space Research, Sorbonnelaan 2, NL-3584 CA
 Utrecht, The Netherlands
\and
Istituto Tecnologie e Studio Radiazioni Extraterrestri, CNR, via
 Gobetti 101, I-40129 Bologna, Italy
\and
Istituto di Fisica Cosmica ed Applicazioni Informatiche, CNR, Via
 U. La Malfa 153, I-90146 Palermo, Italy 
\and
Istituto di Astrofisica Spaziale, Via E. Fermi 21, I-00044 Frascati,
 Italy 
}

\offprints{F. Favata (fabio.favata@astro.estec.esa.nl)}

\date{Received date ; accepted date}

\maketitle 
%\markboth{F. Favata et al., \feh\ in a volume-limited sample of
%  solar-type stars}{\feh\ in a volume-limited sample of
%  solar-type stars}
\begin{abstract}

  We present a broad band X-ray observation of the Cas~A supernova
  remnant, obtained with the 4 narrow field instruments on board the
  BeppoSAX satellite. The X-ray spectrum thus obtained spans more than
  two decades in energy, from $\simeq 0.5$ to $\simeq 80$\,keV. The
  complete spectrum is fit with a two-component non-equilibrium
  ionization (NEI) model plus a power-law component which dominates at
  the higher energies. The influence of the hard X-ray tail on the
  parameters derived for the thermal emission is discussed.


\end{abstract}

\section{Introduction}
\label{sec:intro}

The remnants of many supernova explosions are bright X-ray sources,
with their soft X-ray spectra showing strong metal emission lines, and
characteristic temperatures ranging from below 1\,keV up to a few keV.
Thanks to their high X-ray luminosity and their complex spectra, they
have been a frequent target for spectroscopic X-ray observations.  Of
particular interest are the remnants in which the ejected remains of
the parent supernova are visible in the optical and are recognizable
from their very high metal abundance, a class for which the Cas~A
remnant is often considered the prototype. X-ray emission is in these
systems modeled as being due to two distinct plasma components, i.e.\ 
the ejecta from the supernova explosion, heated to X-ray temperatures
by the reverse shock, and the shocked surrounding circumstellar medium
(CSM), heated by the forward shock ploughing through it. In this
simplified view, the two components are expected to have different
characteristics, with the ejecta very metal rich, and with abundance
values reflecting the results of the nucleosynthesis which took place
in the supernova progenitor as well as in the explosion itself. At the
same time, the emission from the shocked CSM should show abundance
values resulting from a mix of the interstellar medium and of the
stellar wind from the progenitor.  Analysis of the X-ray spectrum can
thus be used to derive an estimate of the mass of the shocked
material, for both emission components, as well as the metal abundance
and the mass fraction of each element in the ejecta. These parameters
can be used to directly verify models of nucleosynthesis in supernova
progenitors.

\begin{figure*}[htbp]
  \begin{center}
    \leavevmode
%    \picplace{7.0 cm}
    \epsfig{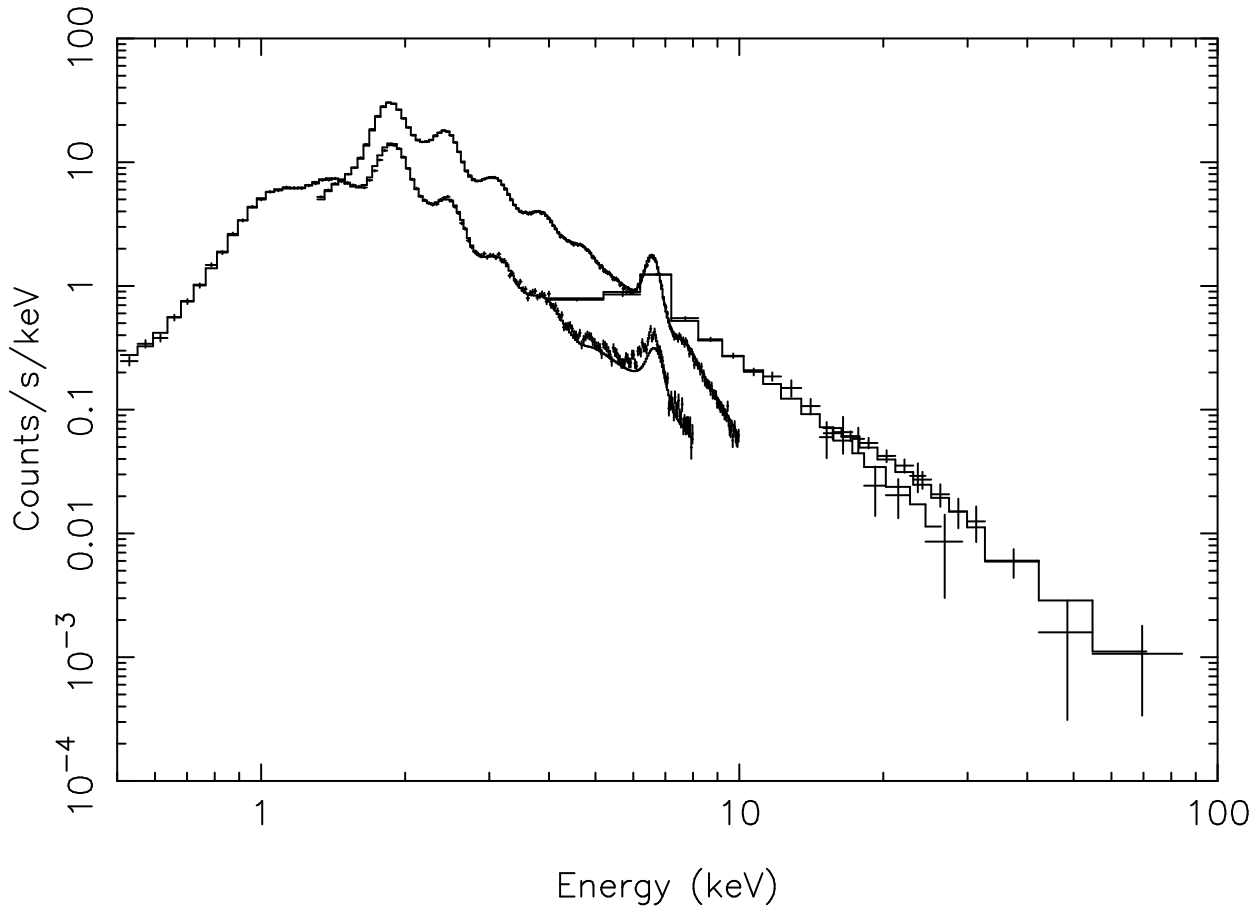}
  \end{center}
  \caption{
    The observed composite BeppoSAX spectrum of Cas~A, using data from the
    LECS, MECS, HPGSPC and PDS instruments, together with the best-fit
    two-component NEI plus power-law model.}
  \label{fig:spec}
\end{figure*}

Given the young age of the Cas~A remnant ($\simeq 320\,$yr) the
shocked plasma is not expected to have reached ionization equilibrium,
and indeed it is by now accepted that the X-ray emission of supernova
remnants is better modeled with non-equilibrium ionization (NEI)
plasma emission codes than with codes modeling the emission of a
plasma in collisional equilibrium.

At the same time, hard X-ray emission of non-thermal origin has been
detected in some of these remnants, extending to energies of $\ga
100$\,keV. Such a hard X-ray tail is of course of astrophysical
interest by itself, and it is also important since it contributes to
the soft X-rays, thus biasing the interpretation of the thermal
emission (\cite{tlk+96}).

The X-ray satellite BeppoSAX (\cite{bbp+97}) includes four co-aligned
Narrow Field Instruments, with complementary energy band coverage,
ranging from 0.1 to $\ga 100\,$keV: the Low Energy Concentrator
Spectrometer (LECS, \cite{pmb+97}), the three Medium Energy
Concentrator Spectrometers (MECS, \cite{bcc+97}), the High-Pressure
Gas Scintillation Proportional Counter (HPGSPC, \cite{mgs+97}) and the
Phoswich Detector System (PDS, \cite{fcd+97}).

Given both its astrophysical interest, and its usefulness as a
calibration source, Cas~A was included as a target in the Science
Verification Phase (SVP) of the BeppoSAX program. During the SVP Cas~A
was observed at different times. In the present paper we discuss the
spatially-integrated broad-band spectrum of Cas~A, obtained using all
the narrow-field instruments on board the BeppoSAX satellite. In the
present paper the LECS data have been used to cover the spectral range
0.5--8.0\,keV, the MECS data the range 1.5--10\,keV, the HPGSPC data
the range 5.0--30\,keV and the PDS data the range 20--80\,keV.  While
the LECS is sensitive down to 0.1\,keV, interstellar absorption is
such that no source flux is detected below $\simeq 0.5$\,keV.

\section{Analysis}

The Cas~A remnant is several arcmin in size, and thus is, at the
resolution of the LECS and MECS detectors ($\simeq 1$\,arcmin), an
extended source. In the present paper however only the spatially
integrated spectrum of the remnant is discussed. The data analysis was
performed using the SRON SPEX package, following an approach similar
to the one adopted by \cite*{vkb96} --- VKB in the following --- who
analyzed the Cas~A X-ray spectrum from the ASCA SIS. The source
spectrum was modeled by two distinct NEI components, one representing
the emission from the ejecta, the other from the shocked CSM. In our
model the CSM component is assumed to be the hotter one, and the
emission from the ejecta is modeled as the cooler component of the
spectrum\footnote{An alternative temperature structure has been
  proposed by \cite*{bsb+96}, who attribute the cool thermal emission
  to the CSM and the hot emission to the ejecta.}. In addition, a
third component is used, in the form of a power-law, to model the hard
X-ray emission. The power-law component is assumed to extend across
the whole X-ray spectrum, i.e.  no cut-off was imposed. The distance
to the SNR was assumed to be 3.4\,kpc (\cite{rhf+95}).

Given the current residual absolute calibration uncertainties of the
various BeppoSAX instruments, the relative normalization of the
different data sets has been considered as a free parameter, using the
MECS normalization as reference. In practice the LECS and PDS
normalizations agree well with each other, as do the MECS and HPGSPC
ones. The LECS/PDS normalization is $\simeq 25$\% lower than the
MECS/HPGSPC one, in agreement with the values found for other SVP
sources.

\begin{table}[htbp]
\caption{The best-fit parameters for the adopted two-component NEI
  plus power-law model to the Cas~A BeppoSAX spectrum. The range
  between brackets is the formal 90\% range ($\Delta\cq = 2.71$). The
  entries in the ``VKB'' column are averaged values from the
  analysis of the ASCA SIS spectrum of VKB.}
\label{tab:fitpar}
\begin{flushleft}
%\scriptsize
\begin{tabular}{lrr} \hline \\[-10pt]
Element & \multicolumn{1}{c}{SAX} & VKB  \\
\hline
{\bf Ejecta component} & &  \\
$n_{\rm e}n_{\rm H}V~(10^{62}\,{\rm m}^{-3})$ & 0.45 [0.36--0.64] & 2.5 \\
Post-shock $T$ (keV) & 1.25 [1.20--1.33] & 0.7\\
$n_{\rm e}t~(10^{15}\,{\rm m}^{-3}{\rm s})$ & 62 [57--68] & 170\\[0.3pt]
{\bf CSM component} & & \\
$n_{\rm e}n_{\rm H}V~(10^{62}\,{\rm m}^{-3})$ & 83 [68--93] & 270\\
Post-shock $T$ (keV) & 3.8 [3.5--4.2] & 4.2 \\
$n_{\rm e}t~(10^{15}\,{\rm m}^{-3}{\rm s})$ & 188 [165--220] & 130 \\
global abundance & 9.6 [7.5--12] & 3.5\\[0.3pt]
{\bf Power-law component} & & \\
Norm. ($10^{44}$ ph\,s$^{-1}$\,keV$^{-1}$ at 1\,keV) & 9.7 [8.8--10.7] & -- \\
Photon index & 2.95 [2.90--3.04] & -- \\[0.3pt]
{\bf Interstellar absorption} & & \\
\nh\ ($10^{22}$ cm$^{-2}$) & 1.22 [1.19--1.26] & 1.5 \\
\hline
\end{tabular}
\end{flushleft}
\end{table}

\begin{table}[htbp]
\caption{The mass fraction, relative to oxygen, implied by the
  best-fit NEI model for the ejecta. The range indicated within the
  brackets is the formal 90\% range (i.e.\ $\Delta\cq = 2.71$). The
  entries in the ``VKB'' column are averaged values from the
  analysis of the ASCA SIS spectrum of VKB, while the ones in the
  ``JY84'' column are the theoretical calculations from Johnston \&
  Yahil (1984).}
\label{tab:mass}
\begin{flushleft}
%\scriptsize
\begin{tabular}{llll} \hline \\[-10pt]
Element & \multicolumn{1}{c}{SAX} & VKB & JY84 \\
\hline
Ne   & 0.000 [0.000--0.010] & 0.02 & 0.327 \\
Mg   & 0.015 [0.011--0.019] & 0.006 & 0.077 \\
Si   & 0.100 [0.070--0.130] & 0.04 & 0.092 \\
S    & 0.069 [0.046--0.079] & 0.03 & 0.036 \\
Ar   & 0.020 [0.014--0.026] & 0.01 & 0.006 \\
Ca   & 0.020 [0.013--0.024] & 0.01 & 0.006 \\
Fe   & 0.025 [0.021--0.033] & 0.02 & 0.15 \\
Ni   & 0.026 [0.021--0.031] & 0.003 & -- \\
\end{tabular}
\end{flushleft}
\end{table}

Given the complexity of the model we have used and thus the large
number of parameters, not all the parameters can be derived from the
data by a blind fitting procedure. Some choices have been made a
priori based on the current physical understanding of the source.  The
parent supernova is thought to be a helium star, which had shed all of
its hydrogen envelope, and indeed the optical spectrum of the ejecta
indicates that they are essentially devoid of hydrogen and rich in
oxygen. We simulate this, in our ejecta model, by imposing a very
high, fixed oxygen abundance relative to hydrogen, and thus
determining the abundance of the other elements {\em relative to O}.
The O abundance has been fixed to 10$^4$ times solar, but the results
are not sensitive to the precise value used (i.e. a value of 10$^2$ or
10$^3$ would yield very similar results).  For the ejecta component,
in addition to the normalization, the post-shock temperature and the
ionization time, the abundance of all individual elements (Ne, Mg, Si,
S, Ar, Ca, Fe, Ni) is initially left free to vary. One additional
parameter is the abundance of the lighter elements relative to oxygen.
He, C and N have no measurable line emission due to the strong
absorption, yet the value of their abundance (specially C) in the
ejecta is important because they contribute to the continuum emission.
Thus, more carbon yields a higher continuum and thus lower abundance
for the other elements. Therefore it is necessary to fix the He, C and
N abundance for the fit to some ``reasonable'' value.  Our choice has
been to use the He/C and O/C values derived from models of
nucleo-synthesis in supernovae, and in particular the models of
\cite*{jy84}.

\begin{figure}[htbp]
  \begin{center}
    \leavevmode
%    \picplace{7.0 cm}
    \epsfig{file=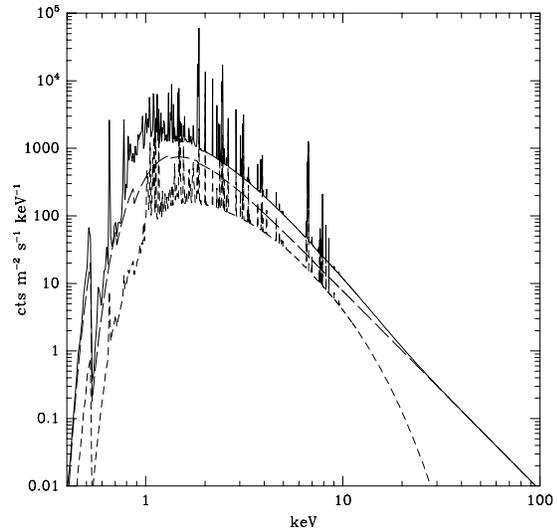, width=7.5cm}
  \end{center}
  \caption{
    The deconvolved broad-band BeppoSAX spectrum of Cas~A, using the data
    from the LECS, MECS, HPGSPC and PDS, together with the individual
    contribution from the best-fit two-component NEI model and the
    power-law high energy tail.}
  \label{fig:decon}
\end{figure}

For the CSM component, in addition to the normalization, the
post-shock temperature and the ionization time, we have left the {\em
  global} abundance free to vary, i.e. we have assumed that the
abundance ratios of elements heavier than N in the shocked CSM are
well modeled by solar values.  The abundance values of He and N were
fixed to 10 times the solar value, an estimate in line with optical
observations of the emission from the shocked CSM (\cite{ck79}). This
implies a depletion of H in the CSM, so that the global abundance of
the CSM is expected to be non-solar due the influence of the strong
stellar wind in the late stages of the progenitor star's life.

\section{Results}

The two-component NEI plus power-law model discussed above provides a
good fit to the observed spectrum across a broad range of energies,
i.e. from 0.5 to 80\,keV, as shown in Fig.~\ref{fig:spec}. Given the
high signal-to-noise ratio of the spectra used, the uncertainties are
likely to be dominated by systematic effects, as for example
uncertainties in the (relative) calibration of the various detectors.
In particular there is some energy-dependent effective-area
inconsistency between the LECS and MECS data (likely linked to the
extended nature of the source), which is evident around the Fe\,K
complex.  Additionally, there is evidence for some residual features
in the HPGSPC spectrum around $\simeq 12$\,keV, which is likely to be
due to residuals in the background subtraction.

The resulting \cq\ of the fit, at 1160 with 382 degrees of freedom is
formally not acceptable. However, given the large residual
systematics, the nominal \cq\ per se cannot be used as a measure of
the acceptability of the model. The best-fit parameters are listed in
Table~\ref{tab:fitpar}, while the derived ejecta mass for each species
is listed in Table~\ref{tab:mass}.

While the fit supplies a good phenomenological description of the data
some of the best fit parameters are rather peculiar. Most striking are
the high Ni abundance and the zero Ne abundance. While Ne has been
found to be under-abundant in the optical (\cite{fes90}) and X-ray
(VKB), the Ne-K lines fall in the same energy range where the rich
Fe-L and Ni-L line complexes are, so that, at the limited energy
resolution of non-dispersive spectrographs, it is difficult to
disentangle real Ne abundance effects from, for example, modeling
problems. The same applies to the Ni abundance, which is derived
exclusively from the L-shell line complex, which is modeled, in SPEX,
somewhat inconsistently with respect to the Fe-L complex.

\section{The non-thermal tail}

The separate contribution of the three components to the Cas~A broad
band X-ray spectrum is shown in Fig.~\ref{fig:decon}. The power-law
component dominates at the higher energies ($\ga 20$\,keV), but it
also gives a sizable contribution to the continuum emission at the
lower energies, being for example comparable in intensity to the
thermal continuum near the Fe\,K complex. Such an additional continuum
clearly influences the derived metal abundances, in particular making
them higher with respect for example to VKB.

Possible explanations for the high-energy tail include (1) synchrotron
radiation from shock-accelerated electrons with energies of several
tens of TeV, similar to the model proposed for SN1006 (\cite{kpg+95};
\cite{rey96}); (2) non-thermal bremsstrahlung from electrons with
energies of several tens of keV, which have just been accelerated from
the tail of the thermal distribution, the so called injection
spectrum; (3) an additional thermal component. Some of our results
depend on the underlying assumptions. Modeling the tail as a power-law
extending down to the lowest energies (as it was done here) is correct
if the tail arises from a physically separate emission component such
as synchrotron radiation. If the emission mechanism is however
non-thermal bremsstrahlung the high energy photons arise from the
non-maxwellian tail of the overall electron distribution, so that the
power-law model will only be a good representation of the emission
above $\simeq$15\,keV. In this case the best fit abundances would be
too high and the estimated normalizations of the thermal
bremsstrahlung too low.  Furthermore, non-thermal electrons can also
contribute to the line emission. Also, possible contribution of
$^{44}$Ti at 68\,keV and 78\,keV (\cite{iea+94}; \cite{tlk+96}) to our
last bin may alter our best fit model.

\section{Discussion}

The model for the X-ray emission of the Cas~A remnant discussed here
follows a similar approach to VKB, with the addition of a power-law
component which dominates at the higher energies.

Some of the VKB ASCA SIS derived parameters are somewhat puzzling,
specially when compared to predictions from theoretical models of
nucleo-synthesis. For example, the ejecta appear to be quite deficient
in their Ne and Mg abundances, with a mass fraction one order of
magnitude lower than expected from nucleo-synthesis models
(\cite{jy84}). The same is true for the Fe abundance, which again is
lower by almost one order of magnitude. 

Assuming that the power-law component (which was not used by VKB in
their analysis) really is a physically separate component of the
spectrum, it significantly influences the mass estimates for the
ejecta, because of differences in the estimated abundances as well as
in the emission measure. For most elements the derived abundance
estimates relative to O increase, when the power-law component is
added, by a factor of about 3, with the exception of Ne and Ni (as
discussed above), with the mass estimates correspondingly decreasing
by $\simeq 40$\% with respect to an analysis using only the two NEI
components.  The apparent lack of Ne and Mg in the ejecta still
stands. The swept-up mass decreases (with respect to VKB) from $\simeq
8\,M_\odot$ to $\simeq 5\,M_\odot$, while the ejecta mass decreases
from $\simeq 4\,M_\odot$ to $\simeq 2\,M_\odot$. The lower derived
ejecta mass does not alter the general conclusion of VKB that the
progenitor of Cas A was a star of low mass (i.e. which had lost most
of its mass).

Some of the assumptions made in the analysis are derived from a priori
considerations about the physics of the remnant, and have therefore to
be critically assessed. Changes in the assumption may lead to changes
in the derived parameters. The two perhaps most critical assumptions
are (1) the temperature structure of the thermal emission components
and (2) the nature and thus precise spectral shape, specially at the
lower energies, of the hard tail component.  The consequence of these
assumptions on the interpretation of the X-ray emission from the
remnant will be examined in a future paper, in which we will also make
use of spatially resolved spectroscopy in order to study if different
components originate from distinct regions.

\acknowledgements{This work was financially supported by NWO. The
  BeppoSAX satellite is a joint Italian and Dutch program.}


\begin{thebibliography}{}

\bibitem[\protect\astroncite{Boella et~al.}{1997a}]{bbp+97}
Boella G., Butler R.~C., Perola G.~C. et~al. 1997a, A\&AS, 122, 299

\bibitem[\protect\astroncite{Boella et~al.}{1997b}]{bcc+97}
Boella G., Chiappetti L., Conti G. et~al. 1997b, A\&AS, 122, 341

\bibitem[\protect\astroncite{Borkowski et~al.}{1996}]{bsb+96}
Borkowski K.~J., Szymkowiak A.~E., Blondin J.~M., Sarazin C.~L. 1996, ApJ, 466,
  866

\bibitem[\protect\astroncite{Chevalier \& Kirshner}{1979}]{ck79}
Chevalier R.~A., Kirshner R.~P. 1979, ApJ, 233, 154

\bibitem[\protect\astroncite{Fesen}{1990}]{fes90}
Fesen R.~A. 1990, AJ, 99, 1904

\bibitem[\protect\astroncite{Frontera et~al.}{1997}]{fcd+97}
Frontera F., Costa E., Dal~Fiume D. et~al. 1997, A\&AS, 122, 371

\bibitem[\protect\astroncite{Iyudin et~al.}{1994}]{iea+94}
Iyudin A. et~al. 1994, A\&A, 284, L1

\bibitem[\protect\astroncite{Johnston \& Yahil}{1984}]{jy84}
Johnston M.~D., Yahil A. 1984, ApJ, 285, 587

\bibitem[\protect\astroncite{Koyama et~al.}{1995}]{kpg+95}
Koyama K., Petre R., Gotthelf E.~V. et~al. 1995, Nat, 378, 255

\bibitem[\protect\astroncite{Manzo et~al.}{1997}]{mgs+97}
Manzo G., Giarrusso S., Santangelo A. et~al. 1997, A\&AS, 122, 357

\bibitem[\protect\astroncite{Parmar et~al.}{1997}]{pmb+97}
Parmar A.~N., Martin D. D.~E., Bavdaz M. et~al. 1997, A\&AS, 122, 309

\bibitem[\protect\astroncite{Reed et~al.}{1995}]{rhf+95}
Reed J.~E., Hester J.~J., Fabian A.~C., Winkler P.~F. 1995, ApJ, 440, 706

\bibitem[\protect\astroncite{Reynolds}{1996}]{rey96}
Reynolds S.~P. 1996, ApJ, 459, L13

\bibitem[\protect\astroncite{The et~al.}{1996}]{tlk+96}
The L.-S., Leising M.~D., Kurfess J.~D. et~al. 1996, A\&AS, 120, 35

\bibitem[\protect\astroncite{Vink et~al.}{1996}]{vkb96}
Vink J., Kaastra J.~S., Bleeker J. A.~M. 1996, A\&A, 307, L41

\end{thebibliography}
\end{document}